\documentclass[twocolumn]{revtex4-2}

\pdfoutput=1
\usepackage[utf8]{inputenc}
\usepackage[english]{babel}
\usepackage[T1]{fontenc}

\usepackage{tikz}
\usepackage{lipsum}

\usepackage{anysize}
\usepackage{algorithmicx,algorithm}
\usepackage{amsfonts}
\usepackage{amssymb,amsmath}
\usepackage{booktabs} 
\usepackage{color}
\usepackage{epsf}
\usepackage{epsfig}
\usepackage{float}
\usepackage{graphicx}
\usepackage{hyperref}
\usepackage{mathrsfs}
\usepackage{microtype}
\usepackage{subfigure}
\usepackage{siunitx}
\usepackage[usenames,dvipsnames]{xcolor}
\usepackage{algpseudocode}

\usepackage{amsmath,amsfonts,amssymb}
\usepackage{mathtools}
\usepackage{bm}
\usepackage{braket}
\usepackage{mathrsfs}
\usepackage{theorem}
\usepackage{multirow}

\newtheorem{theorem}{Theorem}

\newcommand{\dd}{{\rm d}}

\begin{document}


\title{On the Design of Expressive and Trainable Pulse-based Quantum Machine Learning Models}


\author{Han-Xiao Tao}
\affiliation{Center for Intelligent and Networked Systems, Department of Automation, Tsinghua University, Beijing, 100084, China}
\author{Xin Wang}
\affiliation{Center for Intelligent and Networked Systems, Department of Automation, Tsinghua University, Beijing, 100084, China}
\author{Re-Bing Wu}\email[Corresponding author: ]{rbwu@tsinghua.edu.cn}
\affiliation{Center for Intelligent and Networked Systems, Department of Automation, Tsinghua University, Beijing, 100084, China}


\date{\today}

\begin{abstract}
Pulse-based Quantum Machine Learning (QML) has emerged as a novel paradigm in quantum artificial intelligence due to its exceptional hardware efficiency. For practical applications, pulse-based models must be both expressive and trainable. Previous studies suggest that pulse-based models under dynamic symmetry can be effectively trained, thanks to a favorable loss landscape that avoids barren plateaus. However, the resulting uncontrollability may compromise expressivity when the model is inadequately designed. This paper investigates the requirements for pulse-based QML models to be expressive while preserving trainability. We establish a necessary condition pertaining to the system's initial state, the measurement observable, and the underlying dynamical symmetry Lie algebra, supported by numerical simulations. Our findings provide a framework for designing practical pulse-based QML models that balance expressivity and trainability.
\end{abstract}


\maketitle

\section{INTRODUCTION}\label{Sec:introduction}
Quantum computing holds substantial promise for improving the efficiency of both training and inference in machine learning tasks~\cite{arute2019quantum,zhong2020quantum,zhu2022quantum,huang2022quantum}. However, in the noisy intermediate-scale quantum (NISQ) era, the lack of error correction necessitates that quantum machine learning (QML) collaborates with classical computers~\cite{endo2021hybrid,callison2022hybrid} through variational quantum models. In this framework, quantum models execute the inference task, while the classical computer manages the training.

Variational QML models predominantly employ gate-based parameterized quantum circuits, also known as quantum neural networks (QNNs)~\cite{farhi2018classification,mcclean2018barren}. Since all gates are ultimately implemented through control pulses on the relevant physical qubits, recent proposals suggest that QML models may be parameterized directly by these continuous-time control pulses instead of discrete-time gate parameters~\cite{wu2020end,magann2021pulses,choquette2021quantum,meitei2021gate,meirom2023pansatz,liang2022variational,de2023pulse,liang2024napa}. In contrast to gate-based models, pulse-based models are grounded in physical systems without abstraction, making them more accessible to engineers due to their straightforwardness and hardware efficiency~\cite{pan2023experimental,melo2023pulse,ibrahim2022evaluation,egger2023pulse,liang2023towards}.

Within the finite coherence time, pulse-based QML models are more expressive compared to their gate-based counterparts because they function as "infinitely" deep data-re-uploading QNNs~\cite{tao2024unleashing}. We demonstrated this advantage in our previous study~\cite{tao2024unleashing}, where we also proved that pulse-based models achieve full expressivity (i.e., the ability to approximate arbitrary functions~\cite{havlivcek2019supervised,schuld2019quantum,schuld2021supervised,jager2023universal}) when the underlying quantum system is ensemble controllable. This criterion provides a control-oriented framework for designing pulse-based models, analogous to the design of universal quantum computers~\cite{Lloyd1995}. Nonetheless, in systems comprising a large number of qubits, the requirement for controllability becomes excessively stringent as it inevitably leads to barren plateaus within the loss landscape~\cite{larocca2022diagnosing,larocca2025barren}. This phenomenon results in an exponential decay of loss gradients (or, equivalently, the loss variance~\cite{arrasmith2022equivalence, miao2024equivalence}) in the asymptotic temporal limit~\cite{banchi2017driven} and the large qubit-number limit, thus rendering optimization within large-scale systems impracticable~\cite{mcclean2018barren,cerezo2021cost,wang2021noise,arrasmith2021effect,cerezo2021higher,holmes2021barren,sharma2022trainability}. A similar trade-off between expressivity and trainability is observed in gate-based models, where the loss function predominantly concentrates as the circuit attains greater expressivity~\cite{cerezo2021cost,holmes2022connecting,friedrich2023quantum}.

To alleviate the conflict between expressivity and trainability, controllability must be sacrificed in the design of pulse-based models. Larocca \emph{et al.}~\cite{larocca2022diagnosing} revealed that trainability is contingent on the degree of controllability characterized by the size and symmetry of its dynamical Lie algebra. Their findings indicate that a symmetry-restricted model can potentially circumvent barren plateaus, thus providing a practical diagnostic approach for QML model design. Building upon this insight, a Lie-algebraic theory for barren plateaus has been developed, based on which an exact variance formula is derived for the loss function. It is demonstrated that the variance scales inversely with the dimension of the dynamical Lie algebra, provided that either the initial state or observable is contained within the Lie algebra~\cite{ragone2024lie,fontana2024characterizing}. Diaz \emph{et al.}~\cite{diaz2023showcasing} extended beyond the Lie-algebraic framework by analyzing parameterized matchgate circuits. By decomposing the operator space into Lie group modules, they derived an exact variance formula for loss functions applicable to arbitrary initial states and observables, demonstrating that barren plateaus can emerge from generalized global operators and depend on module dimensions rather than solely on the dimension of the dynamical Lie algebra.

The existing literature provides a Lie algebraic framework for evaluating the trainability of uncontrollable pulse-based QML models induced by dynamic symmetry, but the expressivity of these models remains ambiguous. In this study, we demonstrate that an ill-suited selection of either the initial state or the measurement observable can result in the loss of expressivity, and conduct a systematic analysis of the expressivity of pulse-based QML models within the context of dynamic symmetry. In contrast to the Fourier-series analysis typically used to assess the expressivity of data-reuploading gate-based models~\cite{gil2020input,goto2021universal,perez2021one,schuld2021effect,yu2022power}, we utilize the polynomial expansion of nonlinear functions epitomized by the pulse-based model based on Dyson-series analysis (also known as Fliess-series in control theory for input-output analysis of nonlinear systems~\cite{isidori1995nonlinear}). This analysis yields a Lie algebraic tool for designing expressive pulse-based models with dynamic symmetry.

The main contributions of this work are: (i) establishing a necessary condition for expressivity in pulse-based QML models under dynamic symmetry; (ii) demonstrating through numerical simulations that models satisfying this condition can achieve a balance between expressivity and trainability; and (iii) providing a framework for designing practical pulse-based QML models.

The structure of the paper is as follows: Section~\ref{PULSE-BASED QML MODELS} introduces essential background on pulse-based QML models and existing results on trainability under dynamic symmetry. Section~\ref{EXPRESSIVITY OF QML MODELS} analyzes the expressivity of pulse-based models through Dyson-series expansion and derives a necessary condition. In Section~\ref{Balanced trainability and expressivity in multi-qubit QML models}, numerical experiments demonstrate the expressivity of pulse-based models with dynamic symmetry and the balance between expressivity and trainability. Finally, Section~\ref{CONCLUSION AND OUTLOOK} presents conclusions and future perspectives.

\section{PULSE-BASED QML MODELS WITH DYNAMICAL SYMMETRY}
\label{PULSE-BASED QML MODELS}

\begin{figure*}[htbp]
\centering
\includegraphics[width=2.1\columnwidth]{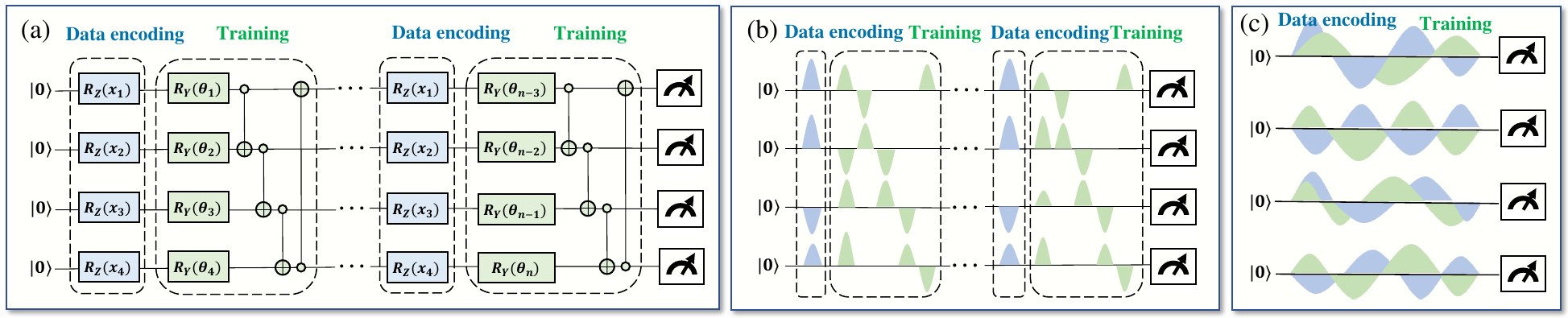}\hfill
\caption{Schematics of (a) a four-qubit gate-based data re-uploading QML model, (b) the pulse-based model compiled from the circuit, and (c) the generalized pulse-based model.}
\label{Fig:pulse-level}
\end{figure*}

We first illustrate the potential advantages of pulse-based QML models through a four-qubit system. In the gate-based data re-uploading model shown in Fig.~\ref{Fig:pulse-level}(a), the input $\mathbf{x}=(x_1,x_2,x_3,x_4)$ is repeatedly uploaded using the encoding circuits $U(\mathbf{x}) =R_{\rm z}(x_1)\oplus \cdots \oplus R_{\rm z}(x_4)$. The trainable circuit blocks comprise $R_{\rm y}(\theta_{k})$ ($k=1,\cdots,n$) rotations about the ${\rm y}$-axis, interspersed with the data-uploading circuit blocks. The final learning outcome is inferred by the expectation-value measurement 
\begin{equation}\label{eq:measurement}
  f(\textbf{x},\Theta)=\langle \psi(\textbf{x},\Theta)|M|\psi(\textbf{x},\Theta)\rangle
\end{equation} 
of a specific observable $M$, where $\ket{\psi(\textbf{x},\Theta)}$ is the output state of the circuit, to approximate multivariable functions by optimizing the parameters $\Theta=\{\theta_{1},\cdots,\theta_n\}$.

Figure~\ref{Fig:pulse-level}(b) illustrates that the physical implementation of the quantum circuit model compiles individual quantum gates into time-varying control pulses on the physical qubits. Let $x_k(t)$ and $\theta_k(t)$ represent the corresponding data-encoding and training control pulses applied to the $k$th qubit. The QML process can be adjusted by $\Theta= \{\theta_1(t),\cdots,\theta_n(t),~t\in[0,T]\}$, resulting in a pulse-based model implemented at the physical level. Furthermore, as shown in Fig.~\ref{Fig:pulse-level}(c), the pulses can be continuously and simultaneously applied without waiting for one another, allowing for compression to reduce inference time. This generalized pulse-based model is free from prior circuit design, features more tunable parameters within the finite coherence time, and can achieve greater expressive power than gate-based models~\cite{tao2024unleashing}.

The pulse-based model implemented with an $n$-qubit quantum device can be characterized as a controlled quantum system:
\begin{equation}
\label{quantum system}
\ket{\dot\psi(t;\textbf{x})}=-i\left[H_{\rm E}(t,\textbf{x})+H_{\rm C}(t)\right]\ket{\psi(t;\textbf{x})},
\end{equation}
where the encoding Hamiltonian
\begin{equation}
    H_{\rm E}(t,\textbf{x})=\sum_{k=1}^mx_k\theta_k(t)H_{k}
\end{equation}
encodes the input variable $\textbf{x}=(x_1,\cdots,x_m)\in\mathcal{X}\subseteq \mathbb{R}^m$ into the dynamical evolution of $\ket{\psi(t;\mathbf{x})}\in\mathbb{C}^{2^n}$. The control Hamiltonian
\begin{equation}
    H_{\rm C}(t)=\sum_{k=m+1}^{m+p}\theta_k(t)H_{k}
\end{equation}
enhances the expressivity of the pulse-based model. The time-variant functions $\theta_1(t),\cdots,\theta_{m+p}(t)$ are control pulses that can be trained to fit the dataset. Similar to gate-based models, the output is inferred through the expectation-value measurement
\begin{equation}\label{eq:measurement_pulse}
  f(\textbf{x},\Theta)=\langle \psi(T;\textbf{x})|M|\psi(T;\textbf{x})\rangle
\end{equation}
with $\Theta=\{\theta_1(t),\cdots,\theta_{m+p}(t),~t\in[0,T]\}$. In practice, certain control pulses may be constant or exhibit limited tunability due to hardware constraints.

Let $\mathfrak{g}=\{iH_1,\cdots,iH_{m+p}\}_{LA}$ be the Lie algebra generated by data-encoding and control Hamiltonians in \eqref{quantum system}, and 
\[
\mathfrak{g} = \mathfrak{c} \oplus \mathfrak{g}_1 \oplus \cdots \oplus \mathfrak{g}_k
\]
be its decomposition as a direct sum of commuting ideals, where $\mathfrak{c}$ is the center of $\mathfrak{g}$ and $\mathfrak{g}_1 \oplus \cdots \oplus \mathfrak{g}_k$ is the semi-simple part. Assuming that either the initial state $\rho=\ket{\psi_0}\bra{\psi_0}$ or the observable $M$ belongs to the Lie algebra $\mathfrak{g}$ (a relaxed condition can be found in~\cite{diaz2023showcasing}), and the QNN is sufficiently deep (i.e., a sufficiently long time duration $T$ for pulse-based QNN) to establish a $2$-design over the Lie group $e^{\mathfrak{g}}$~\cite{ragone2024lie,fontana2024characterizing}), the variance of $f(\textbf{x},\Theta)$ can be estimated as follows~\cite{ragone2024lie,larocca2025barren}
\begin{equation}
\label{eq:variance}
{\rm Var}\left[f(\textbf{x},\Theta)\right] = \sum_{j=1}^k \frac{P_{\mathfrak{g}_j}(\rho)P_{\mathfrak{g}_j}(M)}{\rm dim(\mathfrak{g}_j)},
\end{equation}
where $\rm dim(\mathfrak{g})$ denotes the dimension of the Lie algebra $\mathfrak{g}$, and 
\[
P_\mathfrak{g}(H) \overset{\rm def}{=} \sum_{j=1}^{\rm dim(\mathfrak{g})} \lvert {\rm Tr}(B_j^\dagger H) \rvert^2
\]
denotes the square norm of the orthogonal projection [with respect to the Hilbert-Schmidt inner product $\langle A,B\rangle \overset{\rm def}{=} {\rm Tr}(A^\dagger B)$] of $H$ onto $\mathfrak{g}_\mathbb{C}$ (the complexification of $\mathfrak{g}$). $\{B_j\}^{\rm dim(\mathfrak{g})}_{j=1}$ is an orthonormal basis for $\mathfrak{g}$.

Equation \eqref{eq:variance} indicates that the model becomes untrainable when the underlying control system is fully controllable, because in such cases ${\rm dim}(\mathfrak{g})=4^{n}$ with $\mathfrak{g}=\mathfrak{su}(2^n)$, yielding exponentially decreasing variance as $n$ increases. Therefore, non-trivial dynamic symmetry must be introduced to constrain the quantum state evolution to a manifold whose dimension does not increase exponentially with $n$.

\section{EXPRESSIVITY OF QML MODELS IN PRESENCE OF DYNAMICAL SYMMETRY}
\label{EXPRESSIVITY OF QML MODELS}

The pulse-based model~\eqref{quantum system} is expressive if, for any function $f_0(\textbf{x})$ and a specified error threshold $\epsilon>0$, there exists a suitable control pulse $\Theta$ and a scaling factor $\theta_0$ such that the measurement-induced function \eqref{eq:measurement_pulse} approximates the function with the desired precision, i.e., $\|\theta_0f(\textbf{x},\Theta)-f_0(\textbf{x})\|<\epsilon$. 

In our previous study~\cite{tao2024unleashing}, we demonstrated that a pulse-based model is expressive when the underlying system is ensemble controllable (i.e., any $\ket{\psi(T;\mathbf{x})}$ as a function of $\mathbf{x}$ can be realized within a sufficiently long time $T$). This condition guarantees expressivity for arbitrary selections of the initial state $\ket{\psi_0}$ and the measurement observable $M$. However, as previously noted, this condition is overly stringent for trainability. 

We now demonstrate that uncontrollable pulse-based models can be expressive when the initial state and measurement observable are appropriately selected. To derive the required condition, we expand the output function~\eqref{eq:measurement_pulse} into a polynomial series of $\mathbf{x}$. For convenience, we first rewrite \eqref{quantum system} as a Liouville equation:
\begin{equation}
\dot\rho(t) =\mathcal{L}(t)\rho(t),\quad \rho(0)=\ket{\psi_0}\bra{\psi_0},
\end{equation}
where $\rho(t)=\ket{\psi(t)}\bra{\psi(t)}$ is the density matrix of the system, and
\begin{equation}\label{eq:Liouville}
\mathcal{L}(t)=\sum_{j=1}^{m}x_j\theta_j(t)\mathcal{L}_{j}+\sum_{j=m+1}^{m+p}\theta_{j}(t)\mathcal{L}_{j},
\end{equation}
with the Liouvillian defined as $\mathcal{L}_{j}X=-i[H_j,X]$ for $j=1,\cdots,m+p$.

We apply the Dyson series expansion (also known as the Fliess expansion in control theory literature~\cite{isidori1995nonlinear}):
\begin{eqnarray*}
  \rho(t)&=&\rho(0)+\int_0^t \mathcal{L}(t_1)\rho(0){\rm d}t_1\\
&&  +\int_0^t \mathcal{L}(t_1){\rm d}t_1\int_0^{t_1} \mathcal{L}(t_2)\rho(0){\rm d}t_2+\cdots.
\end{eqnarray*}
Substituting \eqref{eq:Liouville} and expanding the series into polynomial terms of $x_1,\cdots,x_m$, we obtain the polynomial expansion:
\begin{eqnarray*}
\label{power series}
f(\mathbf{x},\Theta)&=& {\rm tr}\left[\rho(T)M\right]\\
&=&\sum_{k_1,\cdots,k_m\geq0} C_{k_1,\cdots,k_m}(\Theta)x_1^{k_1}\cdots x_m^{k_m},  
\end{eqnarray*}
where 
\begin{widetext}
\begin{equation}\label{eq:coefficient}
C_{k_1,\cdots,k_m}(\Theta)=\sum_{n=k_1+\cdots+k_m}^\infty \sum_{(j_1,\cdots,j_n)\in \mathbb{J}_{k_1,\cdots,k_m}}c_{j_1\cdots j_n}(\Theta)\langle\psi_0|\mathcal{L}_{j_1}\cdots\mathcal{L}_{j_n}M|\psi_0\rangle.
\end{equation}
\end{widetext}
The index set $\mathbb{J}_{k_1,\cdots,k_m}$ comprises tuples $(j_1,\cdots,j_n)$ ($n\geq k_1+\cdots+k_m$) wherein each index $j$ in $\{1,\cdots,m\}$ appears $k_j$ times. The coefficients $ c_{j_1\cdots j_n}(\Theta)$ are determined by the control pulses as follows:
\begin{equation}\label{eq:coefficients}
\begin{split}
 c_{j_1\cdots j_n}(\Theta)= &\int_0^T\theta_{j_1}(t_1)\dd t_1\int_0^{t_1}\theta_{j_2}(t_2)\dd t_2 \\
  &\cdots\int_0^{t_{n-1}}\theta_{j_n}(t_n)\dd t_n.
\end{split}
\end{equation}

This polynomial expansion demonstrates the inherent nonlinearity in the model induced by a continuous-limit data-reuploading mechanism~\cite{tao2024unleashing}. The pulse-based model can approximate any continuous function if the coefficients $\{C_{k_1,\cdots,k_m}(\Theta)\}$ in the polynomial series are fully adjustable by tuning the control pulses $\theta_1(t),\cdots,\theta_{m+p}(t)$. According to \eqref{eq:coefficient}, the adjustability of these coefficients depends on both the control pulses, through $c_{j_1\cdots j_n}(\Theta)$ terms, and the initial state and observable through $\langle\psi_0|\mathcal{L}_{j_1}\cdots\mathcal{L}_{j_n}M|\psi_0\rangle$ terms. 

If the coefficients $c_{j_1\cdots j_n}(\Theta)$ were mutually independent and fully adjustable, the model would be expressive provided that, for each $(k_1,\cdots,k_m)$, there exist some $(j_1,\cdots,j_n)\in \mathbb{J}_{k_1,\cdots,k_m}$ such that $\langle\psi_0|\mathcal{L}_{j_1}\cdots\mathcal{L}_{j_n}M|\psi_0\rangle \neq 0$. Unfortunately, the mutual independence of $c_{j_1\cdots j_n}(\Theta)$ does not hold, as demonstrated by the counterexample 
$$c_{j_1j_2}(\Theta)+c_{j_2j_1}(\Theta)=c_{j_1}(\Theta)c_{j_2}(\Theta),\quad \forall~i,j.$$

Nevertheless, we can derive a necessary condition from this analysis. Intuitively, if for a particular monomial $x_1^{k_1}\cdots x_m^{k_m}$, all the coefficients $\langle\psi_0|\mathcal{L}_{j_1}\cdots\mathcal{L}_{j_n}M|\psi_0\rangle$ vanish, then the model cannot represent that monomial regardless of the control pulses chosen. This implies that the model loses expressivity for that monomial. Therefore, to approximate arbitrary functions, the model must be able to represent all monomials, which requires that for every monomial, at least one of the coefficients $\langle\psi_0|\mathcal{L}_{j_1}\cdots\mathcal{L}_{j_n}M|\psi_0\rangle$ is non-zero.

More formally, observing that the coefficient $C_{k_1\cdots k_m}(\Theta)$ corresponding to the monomial $x_1^{k_1}\cdots x_m^{k_m}$ vanishes when $\bra{\psi_0}\mathcal{L}_{j_1}\cdots\mathcal{L}_{j_n}M\ket{\psi_0}=0$ holds for all $(j_1,\cdots,j_n)\in \mathbb{J}_{k_1,\cdots,k_m}$, we conclude that the model loses expressivity under these circumstances, regardless of the adjustability of all $c_{j_1\cdots j_n}(\Theta)$. This leads to the following necessary condition for model expressivity.

\begin{theorem}
\label{thm_1}
Given a sufficiently long duration $T$, the pulse-based QML model~\eqref{quantum system} can approximate any function $f: \mathcal{X}\rightarrow \mathbb{R}$ only when $\bra{\psi_0}\mathcal{L}_{j_1}\cdots\mathcal{L}_{j_n}M\ket{\psi_0}\neq0$ for all $(j_1,\cdots,j_n)\in \mathbb{J}_{k_1,\cdots,k_m}\}$ and all $k_1, \cdots , k_m\geq 0$.
\end{theorem}

To facilitate validation of this condition, we denote by  
\[\mathcal{S}_{k_1,\cdots,k_m}=\{\mathcal{L}_{j_1}\cdots\mathcal{L}_{j_n}M,~(j_1,\cdots,j_n)\in \mathbb{J}_{k_1,\cdots,k_m}\}\]
the set of operators associated with the coefficient of $x_1^{k_1}\cdots x_m^{k_m}$ and introduce a recursive procedure for their evaluation. We first define
\begin{equation}
\begin{split}
\mathfrak{M}(\mathcal{S})=&~{\rm span}\{\mathcal{L}_{\alpha_1}\cdots\mathcal{L}_{\alpha_\ell}X,~X\in\mathcal{S},\\
&\quad m+1\leq \alpha_i\leq m+p,~\ell\in\mathbb{N}\}   
\end{split}
\end{equation}
the submodule generated by the operator set $\mathcal{S}$ with the control Liouvillians. By definition, 
\[\mathcal{S}_{0,\cdots,0}=\mathfrak{M}(M)\]
because $\mathbb{J}_{0,\cdots,0}$ contains only tuples $(j_1\cdots,j_n)$ with $m+1\leq j_k\leq m+p$. 
Starting from $\mathcal{S}_{0,\cdots,0}$, we recursively construct $\mathcal{S}_{k_1,\cdots,k_m}$ as follows:  
\[\mathcal{S}_{k_1,\cdots, k_i,\cdots, k_m}=\mathfrak{M}\left(\mathcal{L}_{i}\mathcal{S}_{k_1,\cdots,k_i-1,\cdots, k_m}\right).\]

Theorem~\ref{thm_1} provides a method for identifying candidate pulse-based QML models that are both trainable and expressive. Although sufficiency for expressivity cannot be rigorously proven, subsequent numerical simulations demonstrate that models meeting this condition tend to be highly expressive. This is attributed to the infinite number of $c_{j_1\cdots j_n}(\Theta)$ terms in \eqref{eq:coefficient}, which offer considerable flexibility even if they are not mutually independent.

\section{NUMERICAL SIMULATIONS}
\label{Balanced trainability and expressivity in multi-qubit QML models}

We demonstrate through numerical simulations the expressivity and trainability of pulse-based QML models under selected dynamic symmetries. 

Given a training dataset $\{(\textbf{x}^{(k)},y^{(k)})\}$ containing $N$ samples, the model is trained by minimizing the loss function
\begin{equation}
\label{cost_func}
L(\Theta,\theta_0)=N^{-1}\sum_{k=1}^N\left\|\theta_0f(\textbf{x}^{(k)},\Theta)-y^{(k)}\right\|^2,
\end{equation}
where $\Theta=\{\theta_1(t),\cdots,\theta_{m+p}(t)\}$. The scaling factor $\theta_0$ adapts the function represented by the model to the range of the dataset. Without loss of generality, the domain of $\textbf{x}$ is set as $\mathcal{X}=[-1,1]^m$~\cite{tao2024unleashing}.

In the numerical simulations, the control pulses are chosen in piecewise-constant form. Each control pulse is divided into $K$ sub-pulses, analogous to the number of layers in gate-based QNN models, with the sub-pulse amplitudes used to parameterize the model. The sampling period is set to $\Delta t=T/K=0.1$, and the model is trained using the Adam optimizer~\cite{Adam} with a learning rate of $0.05$. The simulations are conducted using MATLAB with the QuTiP package~\cite{qutip} for quantum dynamics simulation on a computer equipped with two 10-core Xeon CPUs and 208 GB RAM.

\subsection{Expressivity of pulse-based models with dynamical symmetry}
\label{Approximation of Nonlinear Functions}

For the convenience of model description, we adopt the notation
$$\sigma_\alpha^{(k)}=\mathbb{I}_2\otimes\cdots\otimes\sigma_\alpha\otimes\cdots\otimes \mathbb{I}_2,~~\alpha=\mathrm{x},\mathrm{y},\mathrm{z},$$
for Pauli operators located on the $k$th site in an $n$-qubit system. 

We first demonstrate the approximation of univariate functions using a two-qubit model:
\begin{equation}
\label{eq:simu_1}
H[x;\Theta]=x\sigma_{\rm z}^{(1)}\sigma_{\rm z}^{(2)}+\theta_1(t)\sigma_{\rm x}^{(1)}+\theta_2(t)\sigma_{\rm x}^{(2)}.
\end{equation}
The system is not controllable because it possesses a six-dimensional dynamical symmetry algebra:
\begin{equation}
   \mathfrak{g} =\{\sigma_{\rm z}^{(1)}\sigma_{\rm z}^{(2)},\sigma_{\rm x}^{(1)},\sigma_{\rm x}^{(2)}\}_{LA}=\mathfrak{so}(4)\subset \mathfrak{su}(4).
\end{equation}
We choose the observable $M=\sigma_{\rm z}^{(1)}\sigma_{\rm z}^{(2)}$ and calculate that
\begin{eqnarray*}
\mathcal{S}_{2k}&=&\{\sigma_{\rm z}^{(1)}\sigma_{\rm z}^{(2)},\sigma_{\rm y}^{(1)}\sigma_{\rm z}^{(2)},\sigma_{\rm z}^{(1)}\sigma_{\rm y}^{(2)},\sigma_{\rm y}^{(1)}\sigma_{\rm y}^{(2)}\},\\
\mathcal{S}_{2k+1}&=&\{\sigma_{\rm x}^{(1)},\sigma_{\rm x}^{(2)}\},\quad k=0,1,2,\cdots.
\end{eqnarray*}
 
According to Theorem~\ref{thm_1}, if the system is initialized in the state $|\psi_0\rangle=|0\rangle \otimes |0\rangle$, as is typically done, all odd power terms of $x$ vanish because $\bra{\psi_0}\mathcal{S}_{2k+1} \ket{\psi_0} = \{0\}$ for all $k\in\mathbb{N}$. This implies that such a model can only approximate even functions. However, if the initial state is altered to $|\psi_0\rangle=\left(\frac{2}{\sqrt{5}}|0\rangle+\frac{1}{\sqrt{5}} |1\rangle\right)\otimes|0\rangle$, one can verify that $\bra{\psi_0}\mathcal{S}_k \ket{\psi_0}\neq \{0\}$ holds for all $k\in\mathbb{N}$, suggesting that the model has the potential to approximate any univariate function. 

\begin{figure}[thpb]
\centering
\includegraphics[width=1\columnwidth]{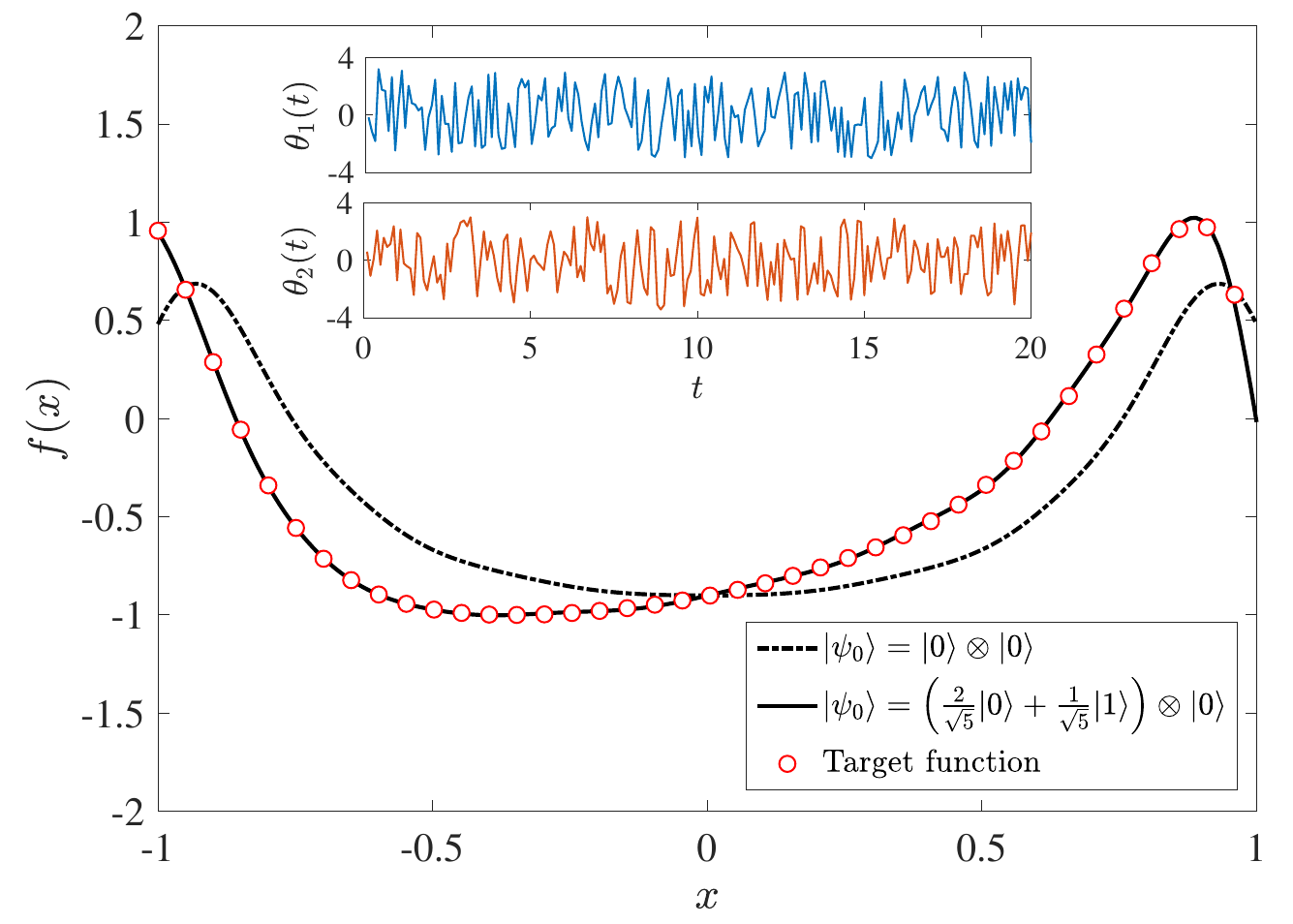}\hfill
\caption{Approximation of the polynomial function \eqref{poly} using the two-qubit pulse-based model~\eqref{eq:simu_1} under $\mathfrak{so}(4)$ dynamic symmetry. The inset displays the trained control pulses.}
\label{Fig:fitting}
\end{figure}

To validate this prediction, we apply model~\eqref{eq:simu_1} to approximate the randomly selected target function:
\begin{eqnarray}
  f_0(x)&=&2x+3x^2+x^3+10x^6 \nonumber \\
  &&+8x^7-3x^9+5x^{10}-13x^{12},  \label{poly}
\end{eqnarray}
from which $200$ points are evenly sampled across $\mathcal{X}=[-1,1]$ to train the model with pulse duration $T=20$ (corresponding to $K=200$). The simulation results presented in Fig.~\ref{Fig:fitting} indicate that the trained QML model accurately fits the target function when $|\psi_0\rangle=\left(\frac{2}{\sqrt{5}}|0\rangle+\frac{1}{\sqrt{5}} |1\rangle\right)\otimes|0\rangle$. However, when the initial state is $|\psi_0\rangle=|0\rangle \otimes |0\rangle$, the model can only approximate the even part of $f_0(x)$ due to violation of the necessary condition in Theorem~\ref{thm_1}.

We also tested the fitting of other randomly selected functions, which can all be approximated by the model, provided that the pulse duration is sufficiently long. The required pulse durations tend to be longer when the target function exhibits greater curvature or includes many higher-order power terms.

We further examine the approximation capability of pulse-based models for more complex bivariate functions using the Hamiltonian:
\begin{equation}
\label{eq:example_bivariate}
\begin{split}
H[\textbf{x};\Theta]=&x_1\theta_1(t)\sigma_{\rm y}^{(1)}\sigma_{\rm y}^{(2)}+x_2\theta_2(t)\sigma_{\rm z}^{(1)}\sigma_{\rm z}^{(2)} \\
&+\theta_3(t)\sigma_{\rm x}^{(1)}+\theta_4(t)\sigma_{\rm x}^{(2)},
\end{split}
\end{equation}
which is similar to \eqref{eq:simu_1} with an additional Hamiltonian encoding $\sigma_{\rm y}^{(1)}\sigma_{\rm y}^{(2)}$, and the observable is chosen as $M=\sigma_{\rm z}^{(1)}+\sigma_{\rm z}^{(2)}$. This model also exhibits $\mathfrak{so}(4)$ dynamic symmetry, and we compute that
\begin{equation}
\mathcal{S}_{k_1,k_2}=\{\sigma_{\rm y}^{(1)},\sigma_{\rm y}^{(2)},\sigma_{\rm z}^{(1)},\sigma_{\rm z}^{(2)}\}
\end{equation}
for all cases where $k_1+k_2$ is even, and
\begin{equation}
\mathcal{S}_{k_1,k_2}=\{\sigma_{\rm x}^{(1)}\sigma_{\rm y}^{(2)},\sigma_{\rm y}^{(1)}\sigma_{\rm x}^{(2)},\sigma_{\rm x}^{(1)}\sigma_{\rm z}^{(2)},\sigma_{\rm z}^{(1)}\sigma_{\rm x}^{(2)}\}
\end{equation}
for all cases where $k_1+k_2$ is odd. Accordingly, the initial state is designated as $|\psi_0\rangle=\frac{1}{\sqrt{2}}|0\rangle\otimes\left(|0\rangle+|1\rangle\right)$, which satisfies the necessary condition in Theorem~\ref{thm_1}. 

In the simulation, the target function is selected as
\begin{equation}
\label{bivariate}
  f_0(x_1,x_2)=(x_1^2+x_2-3)^2+(x_1+x_2^2-1)^2  
\end{equation}
with $50\times 50=2500$ data points evenly sampled across the domain $\mathcal{X}=[-1,1]\times[-1,1]$ and the pulse duration $T$ increased from $0.5$ to $4$. 

\begin{figure*}
\centering
\includegraphics[width=2\columnwidth]{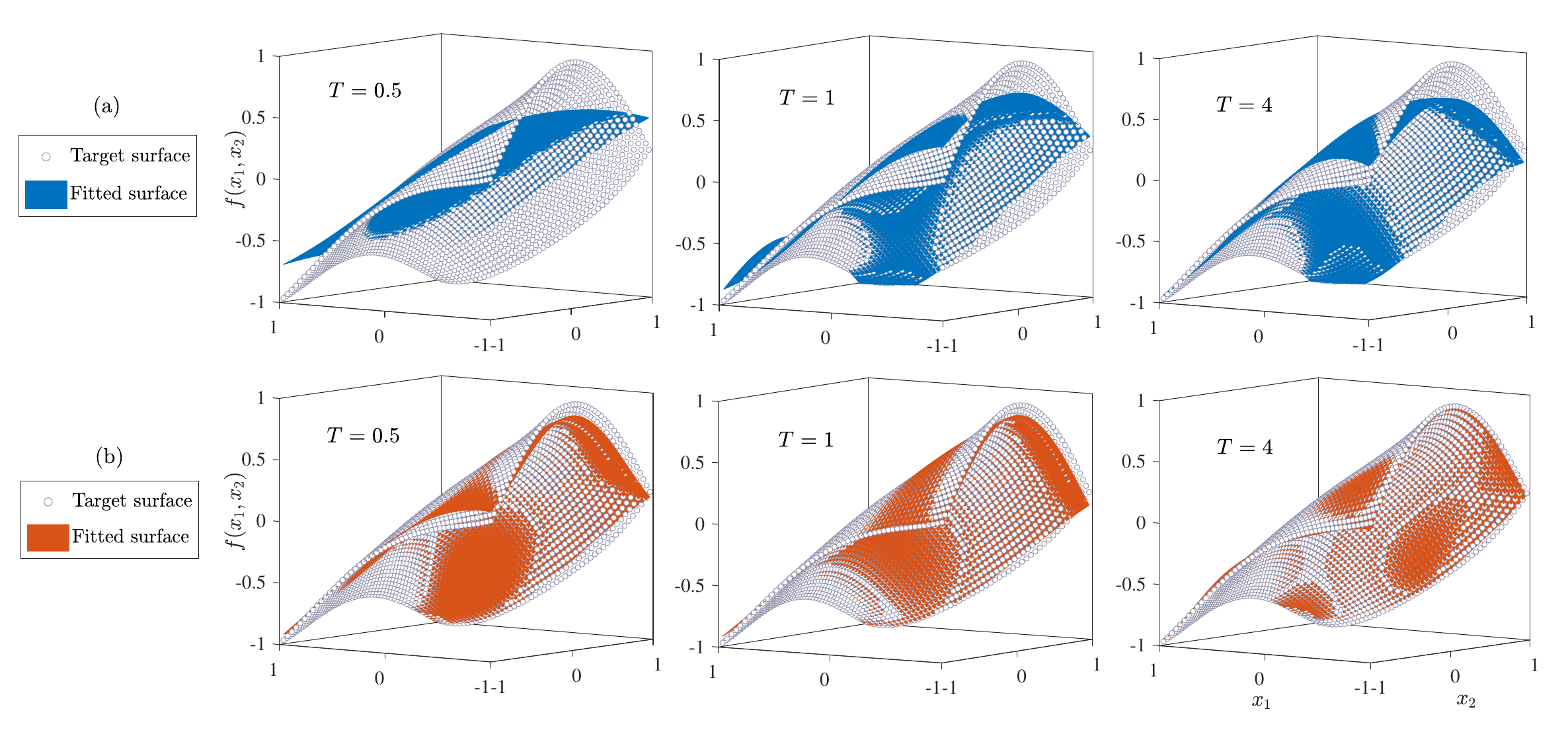}\hfill
\caption{Approximation results of bivariate function by two-qubit models with (a) $\theta_1(t)=\theta_2(t)=1$ and (b) $\theta_1(t)$ and $\theta_2(t)$ freely tunable.}
\label{Fig:bivariate}
\end{figure*}

For comparison, we train two pulse-based models. In the first case, we set the control pulses $\theta_1(t)=\theta_2(t)\equiv1$, which considerably limits the tunability of the coefficients $C_{k_1k_2}(\Theta)$. As illustrated in Fig.~\ref{Fig:bivariate}(a), the fitting progressively aligns with the surface of the target bivariate function as $T$ increases; however, the precision remains limited regardless of how long $T$ is extended. In the second case, we allow $\theta_1(t)$ and $\theta_2(t)$ to vary to enhance the tunability of coefficients $C_{k_1k_2}(\Theta)$, thereby increasing expressivity compared to the first case. The fitting results in Fig.~\ref{Fig:bivariate}(b) confirm this expectation, as they nearly perfectly match the target surface when $T=4$, with a fitting error of $L(\Theta)\leq 10^{-3}$. 

\subsection{Balance between expressivity and trainability}

These simulations demonstrate that pulse-based models can be expressive when uncontrollable due to dynamic symmetry. This establishes a framework for designing large-scale pulse-based QML models that are both expressive and trainable. We select four classes of multi-qubit pulse-based models exhibiting various dynamic symmetries. 

\textit{Model 1.} 
The first model class exhibits ${\mathfrak{su}}(2)$ dynamic symmetry with Hamiltonian:
\begin{equation}
\label{eq:example_su2}
  H_1[x;\Theta]= xJ_{\rm z}+\theta_1(t)J_{\rm x}+\theta_2(t)J_{\rm y}
\end{equation}
with two control pulses, where 
$$\mathfrak{g}=\{iJ_{\rm x},iJ_{\rm y},iJ_{\rm z}\}_{LA}={\mathfrak{su}}(2)$$ 
is the $2^n$-dimensional unitary irreducible representation of ${\mathfrak{su}}(2)$ (see Appendix for detailed description). The observable is chosen as $M=\frac{2}{2^n-1}J_{\rm z}$ and the initial state as $|\psi_0\rangle=|0\rangle^{\otimes n}$. 

\textit{Model 2.} The second model class exhibits ${\mathfrak{su}}(2)^{\oplus n}$ dynamic symmetry with a non-interacting $n$-qubit Hamiltonian:
\begin{equation}
\label{example_2}
  H_2[x;\Theta]= x\sum_{k=1}^n\sigma_{\rm z}^{(k)}+\sum_{k=1}^n \left[\theta_x^{(k)}(t)\sigma_{\rm x}^{(k)}+\theta_y^{(k)}(t)\sigma_{\rm y}^{(k)}\right]
\end{equation}
with $2n$ control pulses. The observable is $M=\frac{1}{n}\sum_{k=1}^n\sigma_{\rm z}^{(k)}$, and the initial state $|\psi_0\rangle=|0\rangle^{\otimes n}$. 

\textit{Model 3.} The third model class, selected from Ref.~\cite{wiersema2024classification}, exhibits ${\mathfrak{so}}(n)$ dynamic symmetry with Hamiltonian:
\begin{equation}
\label{example_3}
  H_3[x;\Theta]= x\sigma_{\rm x}^{(1)}\sigma_{\rm y}^{(2)}+\sum_{k=1}^{n-1} \theta_k(t)\sigma_{\rm x}^{(k)}\sigma_{\rm y}^{(k+1)}
\end{equation}
with $n-1$ control pulses. The observable $M=\sigma_{\rm x}^{(n-1)}\sigma_{\rm y}^{(n)}$ and the initial state $|\psi_0\rangle=\frac{1}{2}|0\rangle^{\otimes (n-2)}\otimes(|0\rangle+|1\rangle)\otimes(|0\rangle+i|1\rangle)$. 

\textit{Model 4.} The fourth model class is fully controllable, utilizing the circularly coupled $n$-qubit system from~\cite{tao2024unleashing}:
\begin{align}\label{eq:simulation}
&H_4[x;\Theta]\nonumber\\
=   & ~x\sum_{k=1}^n\sigma_{\rm z}^{(k)}+\sum_{k=1}^n \left[\theta_x^{(k)}(t)\sigma_{\rm x}^{(k)}
+\theta_y^{(k)}(t)\sigma_{\rm y}^{(k)}\right]\nonumber\\
&+\theta_z^{(1)}(t)\sigma_{\rm z}^{(1)}\sigma_{\rm z}^{(2)}+\cdots+\theta_z^{(n)}(t)\sigma_{\rm z}^{(n)}\sigma_{\rm z}^{(1)}
\end{align}
with $3n$ control pulses. The observable is $M=\sigma_{\rm z}^{(1)}$ and the initial state $|\psi_0\rangle=|0\rangle^{\otimes n}$.

The initial states and measurement observables in the four models all satisfy the necessary condition in Theorem \ref{thm_1} for expressivity. Additionally, the observables are selected from the associated Lie algebra $\mathfrak{g}$ to ensure trainability~\cite{ragone2024lie}. The theoretical variance of the measurement output \eqref{eq:measurement_pulse} is presented in Table~\ref{table_1} (see Appendix for detailed calculations), indicating that the variance is inversely proportional to the dimension of the dynamic Lie algebra $\mathfrak{g}$.

\begin{figure}[thpb]
\centering
\includegraphics[width=1\columnwidth]{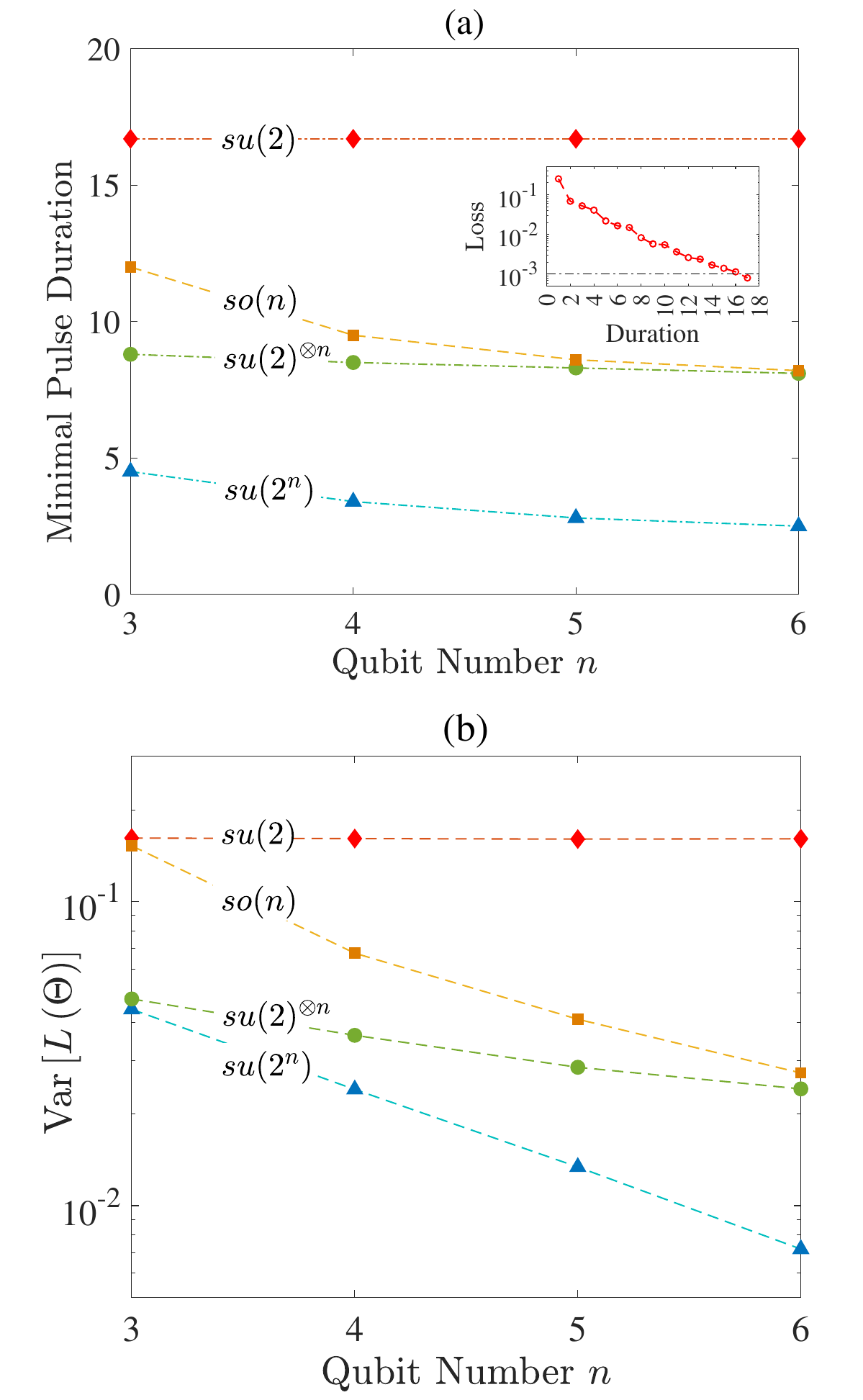}\hfill
\caption{Comparison of expressivity and trainability between model classes under $\mathfrak{su}(2)$, $\mathfrak{su}(2)^{\otimes n}$, $\mathfrak{so}(n)$ and $\mathfrak{su}(2^n)$ dynamic symmetries: (a) Expressivity (evaluated by minimal pulse duration when training loss reaches $10^{-3}$) versus qubit number. (b) Sample variance of loss function versus qubit number. The inset shows how the minimal pulse duration is obtained.}
\label{Fig:T-Loss}
\end{figure}

We trained these models to fit the univariate function~\eqref{poly}. Due to computational resource limitations, we used up to six qubits. As shown in the inset of Fig.~\ref{Fig:T-Loss}(a), each model approximates the target function with arbitrarily small error when the pulse duration $T$ is sufficiently extended. This indicates that model expressivity grows with $T$, providing a measure to quantify expressivity. We adopt the minimal pulse duration required for the training loss to fall below $10^{-3}$ as a measure of expressivity (see inset for illustration).

Using this measure, Fig.~\ref{Fig:T-Loss}(a) compares how expressivity varies with system size across the four model classes. We observe that, except for $\mathfrak{su}(2)$ whose dimensionality is constant, expressivity increases with the number of qubits. Moreover, for sufficiently large numbers of qubits, expressive power increases in the order $\mathfrak{su}(2) < \mathfrak{su}(2)^{\otimes n} < \mathfrak{so}(n) < \mathfrak{su}(2^n)$, indicating that expressivity improves with larger dynamical symmetry Lie algebras.

Regarding model trainability, we estimate the variance of the loss function $L(\Theta)$ numerically. Note that the variance of $L(\Theta)$ differs from ${\rm Var}[f(\mathbf{x},\Theta)]$ in Table~\ref{table_1} since $L(\Theta)$ is a quadratic function of $f(\mathbf{x},\Theta)]$, and no analytic formula is available. Therefore, we randomly select 1000 sets of control pulses to evaluate $L(\Theta)$ and its variance numerically. Since the variance also depends on the pulse duration $T$, we choose $T$ sufficiently long for the estimated variance to converge to a stationary value, corresponding to the model approaching a $2$-design.

Fig.~\ref{Fig:T-Loss}(b) shows the relationship between the stationary variance and the number of qubits. Consistent with the theoretical analysis in Table~\ref{table_1}, the variance for $\mathfrak{su}(2^n)$ decreases exponentially with increasing qubit number $n$, indicating a barren plateau. In contrast, the variance for $\mathfrak{su}(2)$ remains constant due to its constant dimensionality, while decreasing polynomially for $\mathfrak{su}(2)^{\otimes n}$ and $\mathfrak{so}(n)$.

These results demonstrate that barren plateaus can be avoided by restricting dynamical evolution to submanifolds of the quantum state Hilbert space through engineered dynamic symmetry, achieving a favorable balance between expressivity and trainability. We note that overall expressivity and trainability depend strongly on the Hamiltonian structure, initial state, and observable. For example, $\mathfrak{so}(n)$ models with larger $n$ are less expressive than smaller $\mathfrak{su}(2)^{\otimes n}$ models for small qubit numbers, since the latter have more tunable control pulses. However, their scaling behavior matches theoretical predictions, indicating that $\mathfrak{so}(n)$ models become more expressive than $\mathfrak{su}(2)^{\otimes n}$ models for sufficiently large $n$. 

\begin{table}[t]
\begin{center}
\renewcommand{\arraystretch}{2} 
\setlength{\tabcolsep}{3pt}
\caption{Symmetry Lie algebra, dimension, and variance of the inference output~\eqref{eq:measurement_pulse} for each model.}
\label{table_1}
\begin{tabular}{c|c|c|c}
\hline 
Model & $\mathfrak{g}$ & ${\rm dim}( \mathfrak{g})$ & ${\rm Var}\left[f(\textbf{x},\Theta)\right]$ \\
\hline \hline
1 & ${\mathfrak{su}}(2)$ & $3$ & $\frac{1}{3}$ \\
\hline 
2 & ${\mathfrak{su}}(2)^{\oplus n}$ & $3n$ & $\frac{1}{3n}$ \\
\hline 
3 & ${\mathfrak{so}}(n)$ & $\frac{n(n-1)}{2}$ & $\frac{2}{n(n-1)}$ \\
\hline 
4 & ${\mathfrak{su}}(2^n)$ & $4^n-1$ & $\frac{1}{2^n+1}$ \\
\hline 
\end{tabular}
\end{center}
\end{table}

\section{CONCLUSION AND OUTLOOK}
\label{CONCLUSION AND OUTLOOK}

In conclusion, we have established a comprehensive framework for designing practical pulse-based QML models that are both expressive and trainable. Our main contributions include: (i) a necessary condition for expressivity in terms of the initial state, observable, and dynamical Lie algebra; (ii) numerical validation of the condition and its impact on expressivity; and (iii) a demonstration of the balance between expressivity and trainability in various models with dynamic symmetry. Our approach combines Dyson polynomial-series expansion with existing Lie algebraic theories for barren plateaus in quantum systems with dynamic symmetry. Theoretical analysis and numerical simulations demonstrate that dynamic symmetry enables construction of expressive and trainable pulse-based models suitable for hardware-efficient implementation on NISQ devices. 

Whether the proposed necessary condition is sufficient for expressivity of QML models with dynamic symmetry remains an open question. Our simulations indicate that complete expressivity is not always achieved in models satisfying this condition, particularly with constrained control pulses. We conjecture that sufficiency holds when all control pulses are tunable and the pulse duration is sufficiently long, but this requires future verification.

We also note that even if pulse-based models are expressive and trainable, the loss landscape may contain undesirable local minima due to dynamic symmetry, which can trap the training process in suboptimal solutions. Although we rarely encountered this issue in our simulations, this possibility necessitates careful design of training algorithms.

Beyond the trade-off between model expressivity and trainability, an important direction for future work is the trade-off with model generalizability, which concerns model performance on unseen data. Balancing these three attributes has significant practical importance and will be the focus of future work.



%


\appendix

\section{VALIDATION OF EXPRESSIVITY AND VARIANCE CALCULATIONS FOR THE FOUR MODELS}
\subsection{Model 1}
The Hamiltonian is:
\begin{equation}
\label{eq:example_su2_appx}
  H_1[x;\Theta]= xJ_{\rm z}+\theta_1(t)J_{\rm x}+\theta_2(t)J_{\rm y},
\end{equation}
where $\mathfrak{g}=\{iJ_{\rm x},iJ_{\rm y},iJ_{\rm z}\}_{LA}={\mathfrak{su}}(2)$ is the $2^n$-dimensional unitary irreducible representation of ${\mathfrak{su}}(2)$. Let $J_{\pm} = J_{\rm x} \pm i J_{\rm y}$ denote the raising and lowering operators, and 
$$\left\{|s\rangle \mid s = -\frac{d-1}{2}, -\frac{d-3}{2}, \dots, \frac{d-3}{2}, \frac{d-1}{2} \right\}$$
the basis states. In this basis:  
\begin{eqnarray*}
J_z |s\rangle &=& s |s\rangle, \\
J_\pm |s\rangle &=& \sqrt{\left(\frac{d-1}{2} \mp s\right)\left(\frac{d+1}{2}\pm s\right)} |s\pm 1\rangle.
\end{eqnarray*}

Since the maximum and minimum eigenvalues of $J_{\rm z}$ are $\pm \frac{2^n-1}{2}$, and the target function range is $[-1, 1]$, we normalize the observable as $M=\frac{2}{2^n-1}J_{\rm z}$. From Theorem~\ref{thm_1}:
\begin{eqnarray*}
\mathcal{S}_{k}=\{J_{\rm x},J_{\rm y},J_{\rm z}\},k\in N.
\end{eqnarray*}
With initial state $|\psi_0\rangle=|0\rangle^{\otimes n}$, the model satisfies the expressivity condition. 

The model is uncontrollable with $\rm dim(\mathfrak{g})=3$. Using the orthonormal basis 
\[
\left\{\frac{iJ_{\rm x}}{\sqrt{\langle J_{\rm x},J_{\rm x}\rangle}}, \frac{iJ_{\rm y}}{\sqrt{\langle J_{\rm y},J_{\rm y}\rangle}}, \frac{iJ_{\rm z}}{\sqrt{\langle J_{\rm z},J_{\rm z}\rangle}}\right\}
\]
for $\mathfrak{su}(2)$, we obtain:
\begin{eqnarray*}
{\rm Var}\left[f(\textbf{x},\Theta)\right]&=& \frac{1}{3} P_{\mathfrak{g}}(\rho) P_{\mathfrak{g}}(M)\\ 
&=&\frac{1}{3}\left(\frac{2^n-1}{2}\right)^2 \frac{1}{\langle J_{\rm z},J_{\rm z}\rangle}\left(\frac{2}{2^n-1}\right)^2{\langle J_{\rm z},J_{\rm z}\rangle}\\
&=&\frac{1}{3}.
\end{eqnarray*}

\subsection{Model 2}
The Hamiltonian is:
\begin{equation}
\label{example_2_appx}
  H_2[x;\Theta]= x\sum_{k=1}^n\sigma_{\rm z}^{(k)}+\sum_{k=1}^n \left[\theta_x^{(k)}(t)\sigma_{\rm x}^{(k)}+\theta_y^{(k)}(t)\sigma_{\rm y}^{(k)}\right].
\end{equation}
The generated Lie algebra is $\mathfrak{g}= \mathfrak{g}_1 \oplus \cdots \oplus \mathfrak{g}_n$, where each $\mathfrak{g}_j={\mathfrak{su}}(2)$, giving $\dim(\mathfrak{g}) = 3n$. In this case: 
\begin{eqnarray*}
\mathcal{S}_{k}=\{\sigma_{\rm x}^{(i)},\sigma_{\rm y}^{(i)},\sigma_{\rm z}^{(i)}\}_{1 \leq i \leq n} \text{ for } k\in N.
\end{eqnarray*}
We verify that $\bra{\psi_0}\mathcal{S}_k \ket{\psi_0}\neq\{0\}$ for all $k\in N$, satisfying Theorem~\ref{thm_1}.  

The orthonormal basis for $\mathfrak{g}_j$ is $i\{2^{-n/2}\sigma_{\rm x}^{(j)}, 2^{-n/2}\sigma_{\rm y}^{(j)}, 2^{-n/2}\sigma_{\rm z}^{(j)}\}$, and we have $M_{\mathfrak{g}_j}=\frac{1}{n}\sigma_{\rm z}^{(j)}$ and $\rho_{\mathfrak{g}_j}=2^{-n}\sigma_{\rm z}^{(j)}$ for all $j$. Then:
\begin{eqnarray*}
{\rm Var}\left[f(\textbf{x},\Theta)\right]&=& \sum_{j=1}^n \frac{1}{3} P_{\mathfrak{g}_j}(\rho) P_{\mathfrak{g}_j}(M)\\
&=& \sum_{j=1}^n \frac{1}{3} \cdot 2^{-n} \cdot \frac{1}{n^2} \cdot 2^{n}\\
&=& \frac{1}{3n}.
\end{eqnarray*}

\subsection{Model 3}
The Hamiltonian is:
\begin{equation}
\label{example_3_appx}
  H_3[x;\Theta]= x\sigma_{\rm x}^{(1)}\sigma_{\rm y}^{(2)}+\sum_{k=1}^{n-1} \theta_k(t)\sigma_{\rm x}^{(k)}\sigma_{\rm y}^{(k+1)}.
\end{equation}
We derive:
\begin{eqnarray*}
\mathcal{S}_{k}&=&\{\widehat{\sigma_{\rm x}^{(i)}\sigma_{\rm y}^{(j)}}\}_{1 \leq i < j \leq n} \text{ for } k\in N,
\end{eqnarray*}
where $\widehat{\sigma_{\rm x}^{(i)}\sigma_{\rm y}^{(j)}}=\sigma_{\rm x}^{(i)} \sigma_{\rm z}^{(i+1)} \cdots \sigma_{\rm z}^{(j-1)}\sigma_{\rm y}^{(j)}$. With initial state $|\psi_0\rangle=|0\rangle^{\otimes (n-2)}\otimes(\frac{1}{\sqrt{2}}|0\rangle+\frac{1}{\sqrt{2}}|1\rangle)\otimes(\frac{1}{\sqrt{2}}|0\rangle+\frac{i}{\sqrt{2}}|1\rangle)$, we verify that for $n \geq 3$, the model satisfies Theorem~\ref{thm_1}.

The Lie algebra is $\mathfrak{g} = {\mathfrak{so}}(n)$ with dimension $\dim(\mathfrak{g}) = \frac{n(n - 1)}{2} \in \mathcal{O}(\text{poly}(n))$. The orthonormal basis for ${\mathfrak{so}}(n)$ is 
\[
\{i2^{-n/2}\widehat{\sigma_{\rm x}^{(i)} \sigma_{\rm y}^{(j)}} \mid 1 \leq i < j \leq n\},
\]
giving $P_{\mathfrak{g}}(M)=2^n$ and $P_{\mathfrak{g}}(\rho)=2^{-n}$. Then:
\begin{eqnarray*}
{\rm Var}\left[f(\textbf{x},\Theta)\right]&=& \frac{2}{n(n-1)} P_{\mathfrak{g}}(\rho) P_{\mathfrak{g}}(M)\\
&=& \frac{2}{n(n-1)}.
\end{eqnarray*}

\subsection{Model 4}
The model possesses full expressivity according to the criterion in~\cite{tao2024unleashing} since the control Hamiltonians generate the full $\mathfrak{su}(2^n)$ Lie algebra.

For variance calculation, based on~\cite{ragone2024lie}:
\[P_{\mathfrak{g}}(M)  =2^n,\quad
P_{\mathfrak{g}}(\rho) =
1-\frac{1}{2^n},\]
yielding
\[{\rm Var}\left[f(\textbf{x},\Theta)\right]=\frac{1}{2^n+1}.\]

\end{document}